# Day-ahead prediction using time series partitioning with Auto-Regressive model

Dennis Cheruiyot Kiplangat[1,2], G. V. Drisya[1], K. Satheesh Kumar[1*]

[1]Department of Futures Studies, University of Kerala, Thiruvananthapuram, Kerala, India.
[2]Department of Statistics & Actuarial Science, Dedan Kimathi University of Technology, Nyeri, Kenya.
Email: dennis.cheruiyot@dkut.ac.ke , drisyavictoria@gmail.com, *kskumar@keralauniversity.ac.in

*Abstract*— Wind speed forecasting has received a lot of attention in the recent past from researchers due to its enormous benefits in the generation of wind power and distribution. The biggest challenge still remains to be accurate prediction of wind speeds for efficient operation of a wind farm. Wind speed forecasts can be greatly improved by understanding its underlying dynamics. In this paper, we propose a method of time series partitioning where the original 10 minutes wind speed data is converted into a two-dimensional array of order *(N x 144)* where *N* denotes the number of days with *144* the daily *10*-min observations. Upon successful time series partitioning, a point forecast is computed for each of the *144* datasets extracted from the *10* minutes wind speed observations using an Auto-Regressive (AR) process which is then combined together to give the $(N+1)^{st}$ day forecast. The results of the computations show significant improvement in the prediction accuracy when AR model is coupled with time series partitioning.

## I. INTRODUCTION

Wind energy is known globally to be the fastest growing sources of alternative energy. It is clean, green, abundant, environmentally friendly and cheap source of electricity once it is up and operational. By 2050, it is projected that the energy demand could double with the overall global population increase [1]. Each megawatt of generated wind power saves at least 500 kgs of green house gases from being emitted into the atmosphere when compared to power generated from gas or coal, thus making wind power one of the most ecologically friendly sources of energy [2]. According to Global Wind Energy Council (GWEC), there has been a steady growth of wind power generation with the total installed capacity of approximately 318 GW as of year-end 2013 such that if the growth of wind power generation continues at the current rate, it would account for more than 12% of the total energy demands by the year 2020 [3]. From identifying the appropriate sites to environmental strategies and planning, deployment of large scale wind energy presents a wide range of challenges that must be addressed for a more cost effective and reliable wind system. Exploiting opportunities in strategic areas like resource, design, operation, integration, and social and environmental impacts can significantly optimize the performance and cost associated with wind power operation. For a long term R&D, International Energy Agency Wind identified and documented four general research topic to be pursued by the international wind community, namely, characterizing the wind resource, developing next generation wind power technology, wind integration and increasing the social acceptance of wind energy [4]. Wind speed prediction which comes under characterization is one of the important fields of interest since it may help the utility operators to decide how much power to order from traditional sources in order to balance demand and supply at the grid [5]. Wind power is a function of wind speed and prior knowledge of wind speed of the order of a couple of hours to a day ahead are useful in framing a good transmission plan with least possible risk [6]. Apart from transmission planning, integrating highly variable wind power into grid is also associated with some other operational and technical challenges which includes power system operating cost, power quality, power imbalances, and power system dynamics [7] and improvement in prediction accuracy ranging from seconds to a few minutes are useful in addressing these challenges [8]. The literature on the various methods of predicting wind speed can be broadly broken down into two types of models: physical models, such as Numerical Weather Prediction model (NWP), using physical considerations of various atmospheric parameters to obtain the best possible estimate and statistical models using statistical techniques. Although in a stable atmospheric condition physical models are very good in identifying recurring patterns and there by predicting long term future, the computational complexity associated in solving such mathematical models makes them unreliable for short term predictions [9; 10]. The method of persistence, which assumes the future value as the last observed one, is the simplest prediction scheme [11] and mostly regarded as the baseline model when assessing the forecasting performance of a proposed model [12]. The new reference method, a simple modified version of persistence method with a trend towards the mean of the time series is capable to achieve 10% RMS error improvements over persistence [13]. Many statistical models suitable for historical measurements of wind speed are available in literature. The capability of Gaussian distribution in explaining random variations in wind over time explained by [14] is one of the first time series model based papers in this field. In the aim of reducing prediction error combination of different forecast techniques, mixing physical and statistical models or short-term and medium-term models, called hybrid methods are also been experimented by many researchers [15; 16; 12; 17]. Suitability of non-linear models in analyzing dynamical variations of wind resource and improving short term prediction of wind speed has also been studied by [18]. The occurrence of wind is highly uncertain, no single technique can be used universally across all locations and time scales for predicting wind speed and there is always a space for new techniques in the aid of improving prediction accuracy. In this paper, we introduce a new method of prediction where wind speed time series partitioned into a number of time series depending upon length of the prediction time. The method is explained in the following section.

## II. METHODOLOGY

### A. Time Series Partitioning

A function *f(t)* is said to be periodic with period *h* if *f(t+h) = f(t)* for all *t* such that the time series f(t), f(t+h), f(t+2h), …, f(t+2nh) will be a constant time series with simplified predictability. This periodic component can be utilised in improving the prediction accuracy of wind speed time series, thus we investigate the diurnal periodicity present in the wind speed time series. The data considered in this work is the wind speed time series data measured at *80* meters high mast at interval of *10* minutes at location; *Latitude: 34.98420°N* and *Longitude: 104.03971 °W* constituting *144* measurements in a day. We employ continuous wavelet transform to investigate the strength of the diurnal periodicity in the wind speed data. The analysis shows strong diurnal oscillations in the data. Consider the wind speed time series $s_1, s_2, s_3...$ We construct an two dimensional array *A* of order *(N x 144)* as follows;

$$A = \begin{bmatrix} s_1 & s_2 & \cdots & s_{143} & s_{144} \\ s_{145} & s_{146} & \cdots & s_{287} & s_{288} \\ \vdots & \vdots & \ddots & \vdots & \vdots \\ s_{(N-2)*144+1} & s_{(N-2)*144+2} & \cdots & s_{(N-2)*144+143} & s_{(N-2)*144+144} \\ s_{(N-1)*144+1} & s_{(N-1)*144+2} & \cdots & s_{(N-1)*144+143} & s_{(N-1)*144+144} \end{bmatrix}$$

where *N* denotes the number of days, i.e., each row denotes each day's measurements. In order to get prediction of $(N+1)^{st}$ day forecast, a point forecast for each column is done using AR model. Even though wind speed oscillations are not strictly periodic, the diurnal periodicity is very strong in wind speed data. Hence each row in the above array contains unstable periodicities and thus the proposed method utilizes this property to improve the prediction accuracy.

### B. Continuous Wavelet Transform

Most of the time series analysis techniques assume a systematic pattern in the data and the most fundamental pattern in many time series is the frequency of compilation or the periodicity of the data. To identify the cyclical components hidden in the data, spectrum analysis on time series is a good tool since it is capable of exploring the seasonal fluctuations of different lengths. For all possible frequencies the power or strength of each periodic component can also be obtained from the spectrum. For non-stationary data like wind speed time series, amplitude and frequency changes with time and traditional spectral analysis methods like fast-Fourier Transform (FFT) are incapable of providing information about time at which power and frequency variations occurs. Simultaneous analysis of high and low frequency fluctuations in the time series is relevant in these cases and Wavelet Transform is well known technique for the same.

Wavelets are filters used to decompose the original time series data into different sub-bands in terms of time and frequency in order to resolve its fine structures [19]. For the decomposing process Wavelet Transform method uses a small wave called mother wavelet, which is then scaled and shifted along the length of the signal to obtain daughter wavelets. Based on orthogonality we can classify wavelet transform as the continuous wavelet transform: which uses non-orthogonal set of wavelets and the discrete wavelet transform: uses a set of orthogonal wavelets. For transforming the signal into the time-frequency domain one may use a basis function or mother wavelet Ψ in which scaling and shifting is applied. The wavelet at the scale *a* and the position *b* can be expressed as;

$$\psi_{a,b}(t) = \frac{1}{\sqrt{a}} \psi\left(\frac{t-b}{a}\right) \quad \text{where} \quad t \in R, \quad a,b \in R, \quad a > 0$$

In other words, $a$ is the scaling parameter which is dilated or shortened to extract detailed information in the signal and $b$ corresponds to translation parameter which helps in analyzing the signal $f(t)$ around the point $b$. Wavelet transform of a signal $f(t)$ is defined as,

$$W_f(a,b) = \frac{1}{\sqrt{c_\psi}} \int f(t)\psi_{a,b}(t) dt$$

Where * corresponds to complex conjugate and,

$$c_\psi = 2\pi \int \frac{|\hat{\psi}(w)^2|}{w} dw$$

is the admissibility constant which ensures ω the existence of inverse wavelet. The symbol denotes Fourier transform. Now the inverse transform gives the wavelet decomposition of the signal $f(t)$ which is defined as;

$$f(t) = \frac{1}{\sqrt{c_\psi}} \iint W_f(a,b)\psi_a, b(t) \frac{dadb}{a^2}$$

Decomposition of the time series data into several components as a function of both time and frequency aid, not only in quantifying constant periodic components hidden in the data but with information about when they are present.

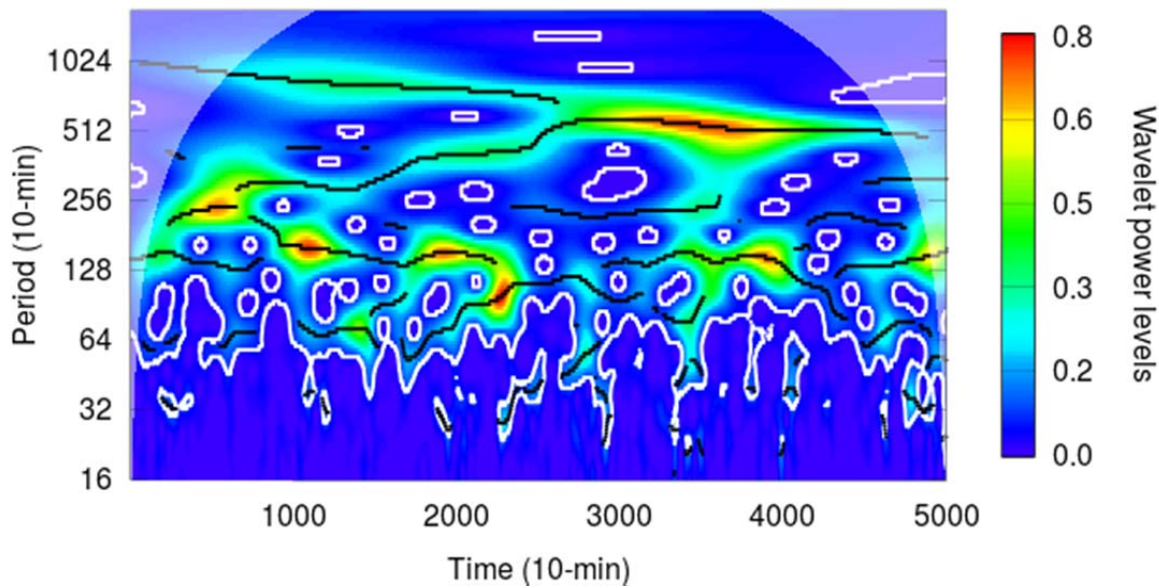

Figure 1**Error! No sequence specified.**: Wavelet power spectrum for the 10-min wind speed data.

*C. Auto-regressive Process*

An auto-regressive (AR) model is commonly used for describing time-varying processes by assuming a linear relationship between output variable and its own historical values. In other words auto-regressive model is just like a multiple regression model where the response variable depends on a set of predictor variables, where the predictor variables are a set of lagged time series. For example, an AR($p$) model is denoted by;

$$X_t = \alpha_1 X_{t-1} + \alpha_2 X_{t-2} + \ldots + \alpha_p X_{t-p} + \varepsilon_t$$

where $\alpha_1, \alpha_2, \ldots, \alpha_p$ are model coefficients, $\varepsilon_t$ is independent and identically distributed random variable or white noise. An optimum choice of model coefficients in AR is important for prediction and while the simplest series AR(1) restricts $\alpha_1$ between -1 and +1, AR(2) has $\alpha_2$ values between -1 and +1 and $\alpha_1 + \alpha_2$ and $\alpha_1 - \alpha_2$ less than 1.

In this paper, the AR coefficients are estimated using the Burg algorithm method which fits an Auto-Regressive model to the input data through minimization of the forward and backward prediction errors while constraining the AR parameters to satisfy the Levinson-Durbin recursion thus providing more stable and robust model parameter estimates [20]. Auto-Regressive models have become more popular for their computational efficiency and simplicity though just like any other stochastic method, the forecast tend towards the mean of the data as prediction period increases.

III. RESULTS AND DISCUSSION

The wind speed data used for analysis are 10-minute resolution data obtained from the National Renewable Energy Laboratory, (http://www.nrel.gov), USA for the period from January 2004 to January 2007 measured at different locations across United States. In this paper, we introduce a new method for day ahead forecasting through the use of time series partitioning discussed in the previous section. The idea is to rearrange the measured wind speed data into several partitions of lag one day and model time series with AR technique to predict one step ahead. Specifications for proper partition of single time series into multiple series can be obtained from the wavelet spectrum which will identify the time of strong periodicity. For the present work we have used Morlet wavelet as the baseline to analyze the frequency structure of time series and the mother Morlet wavelet used for analysis is;

$$\psi(t) = \pi^{\frac{-1}{4}} e^{i\omega t} e^{\frac{-t^2}{2}}$$

Wavelet daughters can be generated as discussed in the previous section and the choice of scale *a* determines the length of the series in the frequency domain that may be covered by wavelet. Usually scale parameter for the wavelet transform is fixed in fractional powers of two. Once the series transformed into wavelet components time-frequency or time-period wavelet energy density can be obtained from the modulus of wavelet transform as

$$\text{Power}(\tau, s) = \frac{1}{s} \cdot |\text{Wave}(\tau, s)|^2$$

and it is called as wavelet power spectrum.

From the wavelet spectrum shown in Figure 1, presence of a strong periodicity associated with the diurnal variation in wind speed is evident at approximately $144^{th}$ period and thus the choice of 144 columns in the partition is made. In other words, since the data of consideration is of 10-min resolution, 144 data points comprise one day and the time series is partitioned into 144 different time series as represented by matrix A in section II. Each column is predicted one step ahead which constitutes a day ahead prediction using AR model. The results have shown that by partitioning the time series, prediction is greatly improved when compared to predictions from a simple AR method. Figures 2 shows one day ahead prediction results obtained from a simple AR method and the proposed time series partitioning method with AR on wind speed data for location, ( *Latitude: 34.98420; Longitude: -104.03971*), along with the observed wind speeds. The performance of simple AR is comparable to the new method initially but its predictions deviate from actual values and level off to the mean value after a short duration. However the proposed method captures the overall underlying dynamics in the data much better.

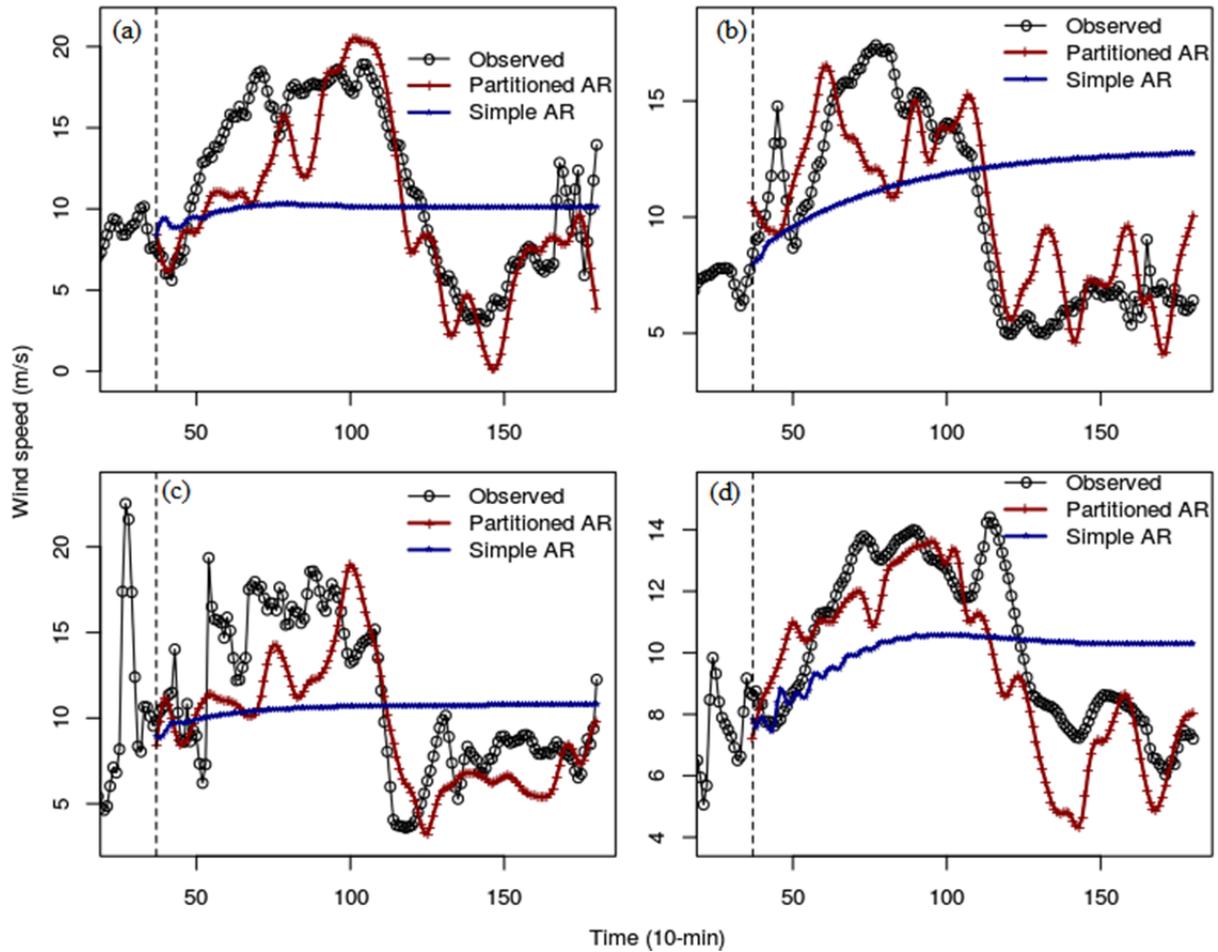

Figure 2: Comparison of predicted values with the actual values for partitioned AR and simple AR at Latitude: 34.98420 and Longitude: -104.03971, for the daily duration from 0000HRS – 2350HRS on (a) 10$^{th}$ May 2004 (b) 9$^{th}$ June 2004 (c) 10$^{th}$ July 2004and ( d) 3$^{rd}$ August 2005.

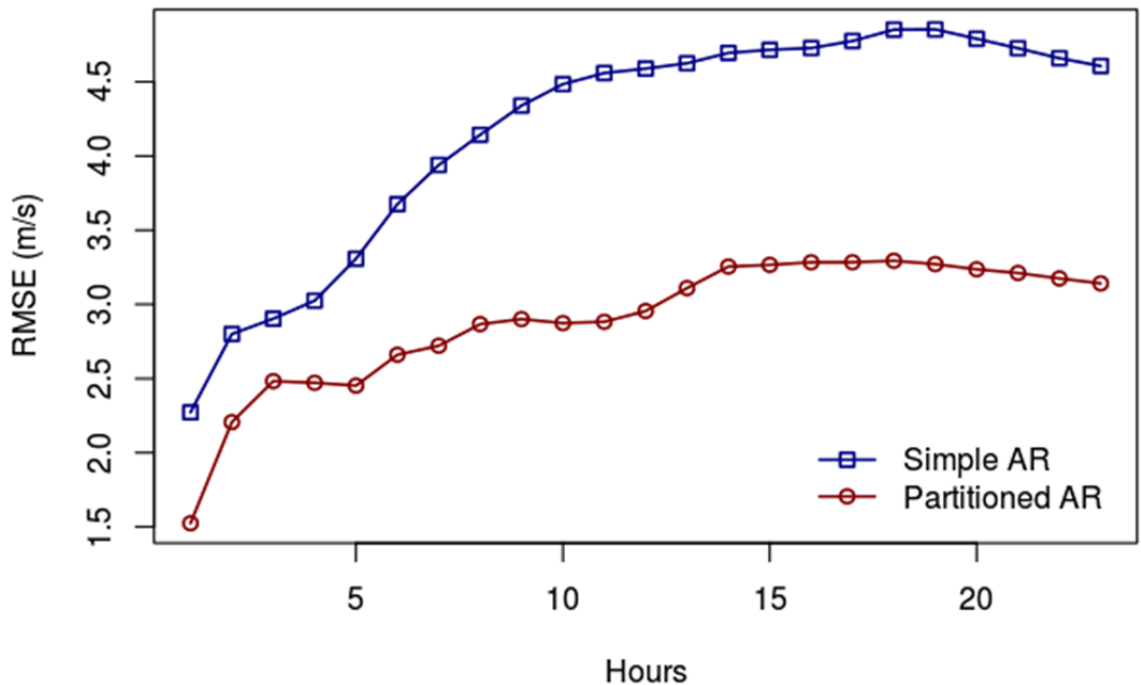

Figure 3: The averaged root mean squared errors of samples shown in Figure 2. for simple AR and partitioned AR predictions.

...

To assess the performance of the proposed method, hourly root mean squared error was computed for the four different time intervals and its averaged root mean squared error is shown in Figure 3 above. It is clear that partitioning the time series and then forecasting using simple AR model outperforms the AR prediction without time series partitioning.

## IV. CONCLUSION

The auto-regressive models have widely been used to model and predict wind speed time series. However the general weakness of the model is that, after faithfully capturing the initial values, it level off to the mean value. In this work we introduce a time series partitioning procedure to improve the accuracy of the prediction. The advantage of the partitioning that we need to predict only one step ahead for each time series to make one day ahead predictions whereas without partitioning one needs to predict 144 step ahead. The results have shown a significant improvement in the prediction accuracy. The proposed method proved to be simple and computationally efficient and it can be used by wind farm operators as an efficient day-ahead prediction tool for wind power generation since accurate speed predictions are critical in grid-connected power distribution and management.

## V. REFERENCES

[1] Geothermal energy: renewable energy and the environment. Glassley, William E. 2014, CRC Press.

[2] Fossil fuel and CO 2 emissions savings on a high renewable electricity system--a single year case study for Ireland. Clancy, JM and Gaffney, F and Deane, JP and Curtis, J and Gallachoir, B O. 2015, Energy Policy, pp. 151--164.

[3] Global Wind Energy Council. Global Wind Energy Council. 2014, Global wind energy outlook.

[4] Long-term research and development needs for wind energy for the time frame 2000--2020. Thor, Sven-Erik and Weis-Taylor, Patricia. 2002, Wind Energy, pp. 73--75.

[5] Powering up with space-time wind forecasting. Hering, Amanda S and Genton, Marc G. 2010, Journal of the American Statistical Association, pp. 92--104.

[6] Wind speed and wind power forecasting using statistical models: Autoregressive moving average (ARMA) and artificial neural networks (ANN). Gomes, Pedro and Castro, Rui. 2012, International Journal of Sustainable Energy Development, pp. 275--293.

[7] Technical challenges associated with the integration of wind power into power systems. Georgilakis, Pavlos S. 2008, Renewable and Sustainable Energy Reviews, pp. 852--863.

[8] Multiscale prediction of wind speed and output power for the wind farm. Wang, Xiaolan and Hui, LI. 2012, Journal of Control Theory and Applications, pp. 251--258.

[9] A Comparison of the impact of QuikScat and WindSat wind vector products on met office analyses and forecasts. Candy, Brett and English, Stephen J and Keogh, Simon J. 2009, IEEE Transactions on Geoscience and Remote Sensing, pp. 1632--1640.

[10] Very short-term wind forecasting for Tasmanian power generation. Potter, Cameron W and Negnevitsky, Michael. 2006, IEEE TRANSACTIONS ON POWER SYSTEMS PWRS, p. 965.

[11] Development and implementation of an advanced control system for medium size wind-diesel systems. Nogaret, E and Stavrakakis, G and Bonin, JC and Kariniotakis, G and Papadias, B and Contaxis, G and Papadopoulos, M and Hatziargyriou, N and Papathanassiou, S and Garopoulos, J and others. 1994, Proceedings of the EWEC, pp. 599--604.

[12] A review of wind power and wind speed forecasting methods with different time horizons. Soman, Saurabh S and Zareipour, Hamidreza and Malik, Om and Mandal, Paras. 2010, North American Power Symposium (NAPS), 2010, pp. 1--8.

[13] A new reference for wind power forecasting. Nielsen, Torben Skov and Joensen, Alfred and Madsen, Henrik and Landberg, Lars and Giebel, Gregor. 1998, Wind energy, pp. 29--34.

[14] Time series models to simulate and forecast wind speed and wind power. Brown, Barbara G and Katz, Richard W and Murphy, Allan H. 1984, Journal of climate and applied meteorology, pp. 1184--1195.

[15] Short-term wind speed forecasting using wavelet transform and support vector machines optimized by genetic algorithm. Liu, Da and Niu, Dongxiao and Wang, Hui and Fan, Leilei. 2014, Renewable Energy, pp. 592--597.

[16] Wind speed forecast model for wind farm based on a hybrid machine learning algorithm. Haque, Ashraf Ul and Mandal, Paras and Meng, Julian and Negnevitsky, Michael. 2015, International Journal of Sustainable Energy, pp. 38--51.

[17] Improved week-ahead predictions of wind speed using simple linear models with wavelet decomposition. Kiplangat, Dennis C and Asokan, K and Kumar, K Satheesh. 2016, Renewable Energy, pp. 38--44.

[18] Deterministic prediction of surface wind speed variations. Drisya, GV and Kiplangat, DC and Asokan, K and Kumar, K Satheesh. 2014, Ann. Geophys.

[19] Wavelet analysis and its applications. Chui, Charles K. 1992, Academic press.

[20] Why Yule-Walker should not be used for autoregressive modelling. De Hoon, MJL and Van der Hagen, THJJ and Schoonewelle, H and Van Dam, H. 1996, Annals of nuclear energy, pp. 1219--1228.

AUTHORS PROFILE

1. D. C. Kiplangat is an Assistant Lecturer at Dedan Kimathi University of Technology, Kenya. He received his MSc. in Applied Statistics from Jomo Kenyatta University of Technology, Kenya and is currently pursuing his PhD in Futures Studies at University of Kerala in the field of wind speed modelling and forecasting.

2. G. V. Drisya is a Ph. D. student in the Department of Futures Studies at the University of Kerala, with particular interests in chaotic time series, wind speed dynamics and Linux shell scripting. She holds a master's degree in Technology Management from University of Kerala and a bachelor's degree in Information Technology from Mahatma Gandhi University. Her research work focus on analysis of deterministic characteristics of wind speed dynamics with a particular attentiveness in short-term prediction.

3. K. Satheesh Kumar received his MSc. in Mathematics from Annamalai University, India in 1988 and his Ph. D. under Faculty of Technology from Cochin University of Science and Technology, India in 1998. Currently, he is an Assistant Professor in the Department of Futures Studies in University of Kerala, India and his research interests include computational modelling and simulations in dynamics and rheology of suspensions and polymer solutions, time series analysis in geophysics and wind dynamics and forecasting.